# Terahertz crystal-field transitions and quasi ferromagnetic magnon excitations in a noncollinear magnet for hybrid spin-wave computation

Gaurav Dubey[1], Brijesh Singh Mehra[1], Sanjeev Kumar[1], Ayyappan Shyam[1], Karan Datt Sharma[1], Megha Vagadia[1], Dhanvir Singh Rana[1]*

[1]*Indian Institute of Science Education and Research Bhopal, 462066, India*

*dsrana@iiserb.ac.in

## ABSTRACT

The complexity of interactions between the crystal-field and unusual non-collinear spin arrangement in non-trivial magnets demands novel tools to unravel the mystery underneath. In this work, we study such interaction dynamics of crystal-field-excitations (CFE) and low-energy magnetic excitations in orthochromite $TmCrO_3$ with controls of temperature and magnetic field using high-resolution magneto-terahertz (THz) time-domain spectroscopy. The THz energy spectrum spanning 0.5-10 meV possesses a low-frequency spin-excitation (magnon) mode and a multitude of CFE modes at 10 K, all of which uniquely embody a range of phenomena. For the magnon mode, a temperature dependence of peak frequency is induced by magnetic interactions between Tm and Cr subsystems. While a change from blue- to red-shift of peak frequency of this mode marks the magnetization reversal transition, the spin reorientation temperature and change of magnetic anisotropy are depicted by different features of field- and temperature-dependent peak frequency dynamics. The modes corresponding to CFE are robust and laden with a multitude of sub-modes which are attributes of non-trivial interactions across different transitions. These modes are suppressed only upon substitution of $Tb^{3+}$ at $Tm^{3+}$ site, which suggests a dominant role of single-ion anisotropy in controlling entire THz excitations spectra. Overall, this remarkable range of phenomena seen through the unique lens of all-optical THz tools provides deeper insights into the origin of magnetic phases in systems with complex interactions between rare-earth and transition metal ions and provides a multitude of a novel combination of closely spaced modes for emerging hybrid spin-wave computation.

# I. INTRODUCTION

The potent combination of rare-earth (RE) ions and transition metal oxides host a range of magnetic phenomena such as negative magnetization, exchange-bias effect, magnetocaloric and magnetoelectric effect in orthochromites (RECrO$_3$), orthoferrites (REFeO$_3$), and manganates (REMnO$_3$), etc, all of which are of fundamental interest while also being technologically relevant for next-generation spintronics, magnonics and memory storage devices. [1–7] The strength and single ion anisotropy of RE moments and their exchange interaction with transition metal magnetic sublattice are the cornerstone of the complex magnetic states of these systems. In systems with large RE magnetic moments, the surrounding crystal field weakly perturbs the multiplet structure generated due to spin-orbit coupling of 4f electrons. This lifts the atomic energy degeneracies; however, the extent of lifting depends on the parity of the number of electrons within the outermost 4f shell by Kramer's theorem. [8] In the case of odd numbers of 4f electrons, the energy levels are split into doubly degenerate states (Kramer's ion), whereas in even electrons, the energy levels are split into singlets/ doublets (non-Kramer's ion). This diversity makes non-Kramer's ion interesting to analyse with the control of thermodynamic variables such as temperature, magnetic field and pressure.

The interaction of 4f electrons with the surrounding crystal field and the exchange field generated due to transition metal ions give rise to electronic and magnetic (spin-gap or spin-wave) excitations. These interactions/excitations, consequently, modifies the magnetic landscape. Conventionally, it requires several levels of experimental investigations to unravel such complexities of the magnetic grounds, namely, the bulk magnetization, neutron diffraction, inelastic neutron scattering, etc. While these tools are sufficient for exploring magnetic phases, they are inadequate to probe the effect of complex magnetic ordering in the electric/dielectric properties. Terahertz (THz) radiation, being uniquely sensitive to both magnetic and dielectric/electric media, can extract significant information of such materials. All the crystal-field and spin/magnon excitations of low energy, falling in THz spectral range, can be probed by contactless and phase-sensitive coherent detection techniques of THz spectroscopy. [9–11] These excitations can be intrinsically coupled to each other linearly / non-linearly either using high intensity THz stimulus or via artificial patterns such as metasurfaces. Such interactions have direct applications in the emergence of quantum technologies requiring dissipation-less and ultrafast data/information processing – two attributes derived from THz magnon (quantized spin-wave) excitations. These magnons can also potentially be hybridized with other THz frequency modes of lattice or crystal-field origin creating new opportunities for hybrid quantum computation. This, however, requires complex magnetic systems which can host a rich wealth of closely spaced excitations with distinct origin and ability for mutual interaction.

Rare-earth orthochromates and orthoferrites are two non-collinear antiferromagnetic series of systems possessing a range of magnetic phenomena relevant to THz science and technology. While orthoferrites have been explored in context and have unravelled a key area of THz magnonics, [12–17] the potential for new phenomena in THz magnetism in their magnetically more complex counterparts, orthochromates, [18] is yet to be realized. $TmCrO_3$ is one such promising system with non-Kramer $Tm^{3+}$ ion in the family of rare earth-based orthochromites. Here, the complex exchange interactions between $Cr^{3+}$ and $Tm^{3+}$ sublattices gives rise to intricate temperature induced magnetism, such as i) at 125 K, $Cr^{3+}$ spins ordered antiferromagnetically with superimposed canting due to the Dzyaloshinkii–Moriya interaction; [1] ii) on lowering of the temperature, the antiparallel coupling of $Cr^{3+}$ and $Tm^{3+}$ sublattices results in compensation of magnetic moments; iii) at 30 K, below which magnetization reversal is observed; iv) on further decreasing the temperature, spins reorient themselves from the c-axis towards a-axis and; [19,20] v) at further lower temperatures (< 2 K) rare-earth ordering takes place. The interplay of Cr-Cr, Tm-Cr, and Tm-Tm interactions makes the ground state complex. Implementation of THz spectroscopy with control of temperature and magnetic field will not only unravel the low-energy magnon/spin-gap, crystal-field and other hybrid modes, but it will also enhance the understanding of evolution of a complex magnetic landscape leading to a variety of THz frequency excitation modes. THz spectroscopy is a proven tool to probe quasiparticles of structural, electrical, magnetic or any other hybrid origin, such as magnons, electromagnons, Higgs mode, crystal-field transition, charge-density-waves, etc, [9,21]

In this work, using magneto-THz spectroscopy, we show a multitudes of THz energy range excitations, namely, the magnon mode and two sets of crystal-field excitations in the magnetic ground state of $TmCrO_3$ along with the magnetic-field control of these modes. The effect of substituting the $Tm^{3+}$ ion with another magnetic ion $Tb^{3+}$ is also studied to understand the role of single ion anisotropy. While these THz modes are relevant in emerging area of low-dissipation magnonic spintronics, the field and temperature dependent dynamics of these excitations unravel so-far illusive new physical phenomena in chromites. We illustrate a "proof of concept" application of hybrid quantum computation devices, which can be realized through the mutual interaction of closely spaced crystal-field and magnon excitations in the narrow frequency range of the THz spectrum in $TmCrO_3$.

## II. EXPERIMENTAL DETAILS

$TmCrO_3$, $Tm_{0.5}Tb_{0.5}CrO_3$ and $TbCrO_3$ polycrystalline samples were prepared by solid-state reaction method. The detailed synthesis process and structural characterization are provided in the supplemental material. In order to investigate the magnetic properties, temperature and magnetic field dependent DC magnetization was performed using SQUID (Quantum Design) magnetometer in the VSM mode for temperatures ranging from 2 to 300 K and in the presence of magnetic field up to 9 T / - 9 T. THz time-domain spectroscopy (TDS) measurements were carried out in transmission mode (Faraday geometry) at varying temperatures (5-300 K) and in magnetic fields (0- 5 T). Dry nitrogen was purged to minimize the water vapor absorption of THz radiation. The THz-Time Domain spectrometer provides time-domain waveforms averaged over 5000 pulses with the fine resolution of 0.014 THz (1 THz = 4.14 meV). Average waveform is then converted into the frequency domain via Fast Fourier Transformation. The output is plotted in terms of absorption coefficient (α) as a function of THz frequency. Absorption coefficient (α) is computed from the relation, $\alpha = \left(\frac{1}{d}\right) ln\ |t|^2$ where t is the ratio of THz transmission with and without the sample and 'd' is thickness of the sample.

## III. RESULTS AND DISCUSSION

### A. Magnetization measurements

Figure 1 (a) shows the temperature dependence of magnetization for $TmCrO_3$ at 50 Oe magnetic field. In $TmCrO_3$, the $Cr^{3+}$ spins order antiferromagnetically at $T_N$ ~ 125 K [Figure 1 (a)]. Around 30 K, the magnetization becomes zero, because the magnetic moments of $Cr^{3+}$- $Cr^{3+}$ ($M_{Cr}$) and $Cr^{3+}$- $Tm^{3+}$ ($M_{Tm}$) interactions become equal and opposite at 30 K. A spin reorientation transition (SRT) below 20 K is depicted by change of slope in dM/dT curve [Inset (a), Figure 1 (a)]. The value of effective magnetic moment obtained by Curie Weiss fit on inverse susceptibility curve is 8.34 $\mu_B$/f.u. [Inset (b), Figure 1 (a)] which is in good agreement with the previously reported values. [22]

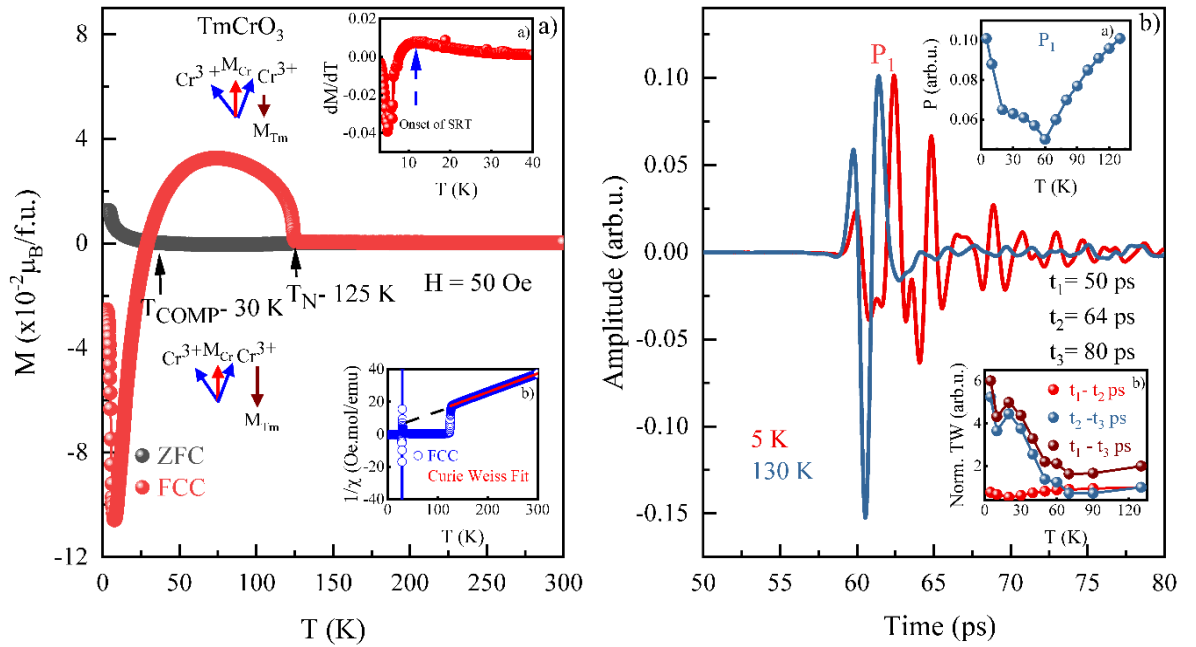

*Figure 1 (a): Magnetization (M) as a function of temperature in ZFC and FCC cycle under an applied magnetic field of 50 Oe of TmCrO$_3$. Inset (a): Temperature derivative of FCC magnetization under magnetic field of 50 Oe, the blue arrow shows onset of spin reorientation transition (SRT). Inset (b): Inverse of magnetic susceptibility with temperature variation, red line show Curie Weiss fit at higher temperature region. 1 (b): Typical transmitted THz waveforms at 5 K and 130 K as a function of time of TmCrO$_3$. Inset (a): Variation of THz peak amplitude (P) of P$_1$ peak denoted in waveform with temperature. Inset (b): Normalised Temporal weight (Norm. TW) transfer with variation in temperature.*

### B. Terahertz measurements

To reveal the microscopic intricacies in this magnetization behaviour we have performed Magneto-THz measurement on TmCrO$_3$. THz waveforms transmitted through TmCrO$_3$ at 130 K (above T$_N$) and 5 K are remarkably different [Figure 1 (b)]. This is evident from both the shape as well as the peak amplitude (P) of P$_1$ peak of these waveforms [Inset (a), Figure 1 (b)]. P$_1$ exhibit non-monotonic behaviour with decreasing temperature [Inset (a), Figure 1 (b)]. With decreasing temperature, the peak position P$_1$ decreases and reaches its minimum at 60 K. A shift in THz waveform can be seen with change in temperature, this temporal shift suggests a change in refractive index, which becomes more pronounced around spin-reorientation transition at 20 K [Figure S7 (a), supplemental material]. These variation in P$_1$ with temperature corresponding to the multiple magnetic transitions in TmCrO$_3$ underline the sensitivity of THz radiation in probing the complex noncollinear magnets. The main THz peak and trailing oscillations depict a dramatic change [Figure 1 (b)]. To systematically unveil them, we defined a term called Normalized THz temporal weight (Norm. TW), which is the integration of THz time domain signal over a finite time interval and normalized with respect to 130 K (>T$_N$) to eliminate the background effect, as

$$THz\ temporal\ weight = \int_{t_1=50}^{t_2=64} E(t).dt + \int_{t_2=64\ ps}^{t_3=80\ ps} E(t).dt \qquad (1)$$

Here, t$_1$ to t$_2$ (50 to 64 ps) encapsulates the main THz peak, whereas t$_2$ to t$_3$ (64 to 80 ps) includes the oscillatory behaviour of the waveform. A maximum transfer of 700 % in spectral weight of THz waveform manifests from lower time (t$_1$- t$_2$) to higher time (t$_2$-t$_3$) scale with the change in temperature [inset (b), Figure 1 (b)].

Detailed mechanism of all magnetic and electronic transitions can be captured in Fourier transform of THz waveforms. Figure 2 (a) shows the THz absorption spectra depicting three major modes denoted by g$_1$, g$_2$ and g$_3$. Mode g$_1$ appears at the lower frequency (0.28 THz at 50 K) which shows a non-monotonous temperature and magnetic field dependence. The broad resonant modes marked as g$_2$ (0.56 THz) and g$_3$ (1.25 THz) appears at around 90 K which split on lowering the temperature. The low frequency mode g$_1$ appears at 60 K and strengthens in magnitude as temperature decreases [Figure 2 (b)]. With lowering temperature, the peak frequency of this mode initially shows blue shift, followed by a red shift at 30 K. This observation is concomitant with onset of magnetization reversal [Figure 1 (a)] and variation of THz peak position $P_1$ [Inset (a), Figure 1 (b)].

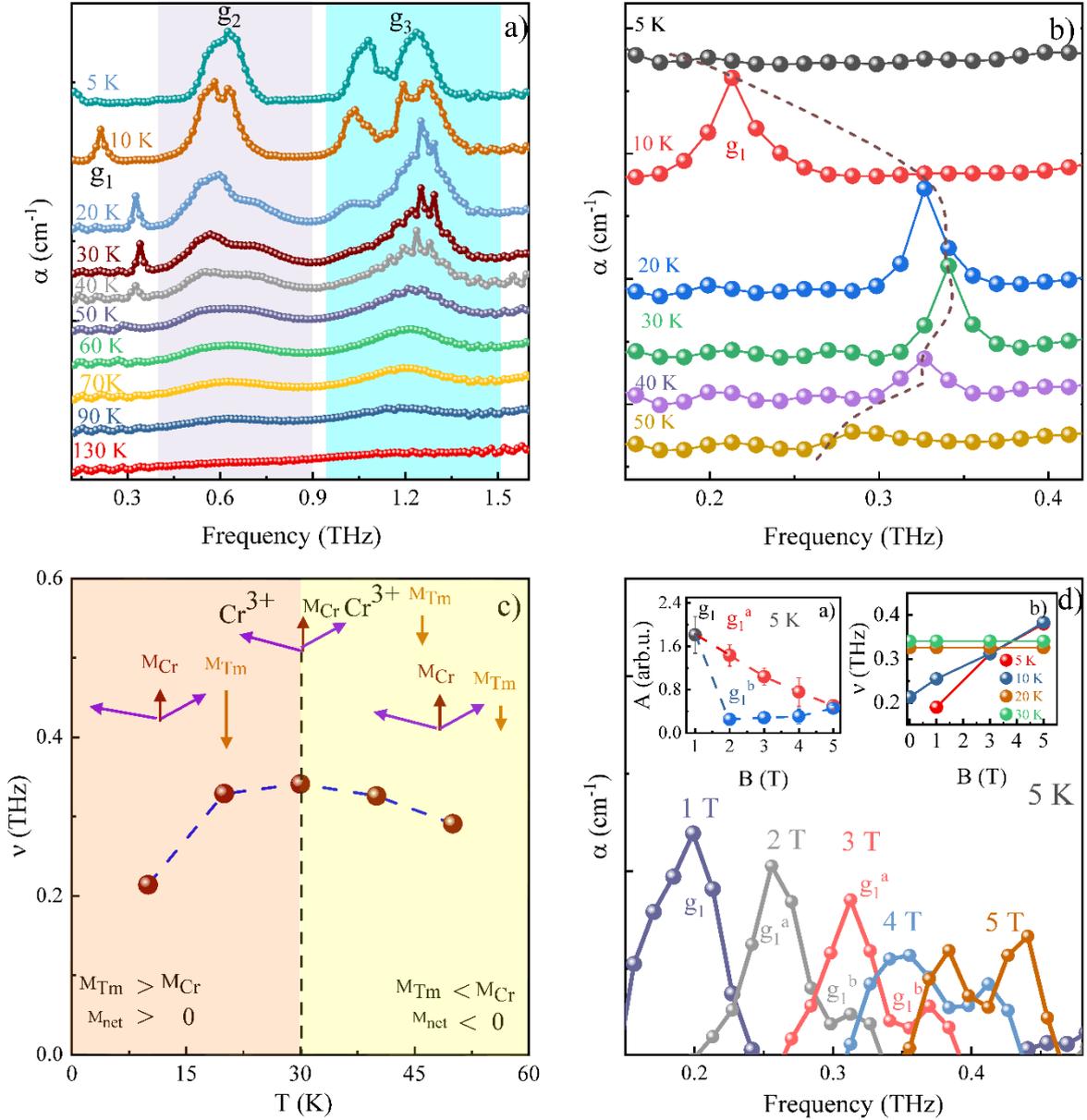

*Figure 2 (a): Absorption coefficient (α) corresponding to different temperatures as a function of THz frequency. (b): Temperature dependence of absorption coefficient (α) of mode $g_1$, the curves have been shifted along the Y axis to enhance visibility in 2 (a) and 2 (b). Dashed line shows the peak frequency shift of $g_1$. (c): Peak frequency (ν) of $g_1$ mode and schematic representing magnetic interaction of $Cr^{3+}$-$Cr^{3+}$ and $Cr^{3+}$-$Tm^{3+}$ sublattices at different temperatures. (d): Magnetic field dependence of absorption coefficient (α) at 5 K. Inset (a): Area under the absorption curve (A) for $g_1$ mode as a function of applied magnetic field at 5 K. Inset (b): Variation of peak frequency (ν) of $g_1$ mode as a function of applied magnetic field corresponding to different temperatures.*

The variation in peak frequency of $g_1$ mode in terms of competition of components of two different magnetic interactions is shown in Figure 2 (c). In TmCrO$_3$, the Tm$^{3+}$ sublattice aligns antiparallel in the negative molecular field produced due to the canted Cr$^{3+}$ spin arrangement. [23] At 30 K, these components become equal and opposite which causes magnetization to become zero, and on lowering the temperature, it becomes negative because of the dominance of $M_{Tm}$ over $M_{Cr}$, which stems from the interactions between Cr$^{3+}$-Tm$^{3+}$ and Cr$^{3+}$- Cr$^{3+}$ sublattices, respectively. The transition in frequency shift of $g_1$ mode at 30 K is associated with the sign change of net magnetic moment. The strength of the $g_1$ mode is related with the magnitude of Tm$^{3+}$ magnetic moment, which increases with a decrease in temperature. It may be noticed that below 30 K the rate of red shift increases and $g_1$ moves out of THz detection range at 5 K. All these relations between the variations of $g_1$ mode and the magnetization data suggest that this mode arises due to interaction between Cr$^{3+}$ and Tm$^{3+}$ sublattices.

To get insights on the magnetic field dynamics of $g_1$, we analysed the THz spectra measured in various applied magnetic fields. In contrast to red shift of $g_1$ upon lowering temperature, a blue shift was observed with increase in magnetic field [Figure 2 (d)]. This is evident from the fact that at 5 K, while this mode is absent in the absence of field, it appears at a finite frequency for a field of 1 T and subsequently shifts towards the higher frequencies with increasing fields. Furthermore, in field of 2 T, this mode splits into two marked as, $g_1^a$ and $g_1^b$ [Figure 2 (d)]. To elucidate the change in $g_1$, area under the absorption curve (A), which is proportional to spectral weight distribution was plotted [Inset (a), Figure 2 (d)]. The area under $g_1^b$ increases, while that of $g_1^a$ decreases with increasing magnetic field. In a field of 5 T, the area of these modes become comparable.

The field induced suppression of the $g_1$ mode is another feature which can be understood in context of magnetization reversal as follows. In the absence of a magnetic field, the Tm$^{3+}$ magnetic sublattice is aligned in the opposite direction to the net magnetic moment of Cr$^{3+}$ sublattices due to the negative internal magnetic field produced by canted Cr$^{3+}$ sublattice. [1,23] In applied field greater than the internal field, the Tm$^{3+}$ magnetic moments tend to align in the direction of the applied field, which disrupts the antiparallel coupling of Cr$^{3+}$ and Tm$^{3+}$ spins and, hence, suppresses the magnetization reversal. [24] Inset (b) of Figure 2 (d) shows the field induced frequency shift of $g_1$ mode at various temperatures. While this shift is dominant at 5 K, it weakens with increasing temperature and then ceases to exist above spin-reorientation temperature. Below 20 K, Cr$^{3+}$ spins orient towards *a*-axis from *c*-axis as per switching of magneto crystalline anisotropy. [19,20] In applied fields at 5 K, the anisotropy of *c*-axis increases which shifts $g_1$ mode to higher THz frequencies which is similar to its temperature-induced variation above 20 K where anisotropy of *c*-axis is greater than *a*-axis, results in blue shift. Along with all these intricate dynamics of $g_1$ with the evolution of different magnetic phases, $g_1$ can be understood to be a quasi-ferromagnetic spin-wave/magnon mode, as reasoned below.

i) In noncollinear antiferromagnets, the spin waves modes are detected in THz spectroscopy because of the interaction of THz magnetic field with net magnetization. Dynamics of these resonant modes can be explained using two-sublattice model.[14,25] The resonant frequency of these quasi-ferromagnetic (qFMR) and quasi-antiferromagnetic (qAFMR) mode depends on magnetic anisotropy constant in the *ac*-plane given as.[7,14]

$$\left(\hbar\omega_{qFMR}\right)^2 = \left[\frac{4E}{(2S)^2}\right][(A_{aa} - A_{cc})\cos 2\theta - 4K_4 \cos 4\theta] \qquad (2)$$

$$\left(\hbar\omega_{qAFMR}\right)^2 = \left[\frac{4E}{(2S)^2}\right][\frac{1}{2}(A_{aa} + A_{cc}) + \frac{1}{2}(A_{aa} - A_{cc})\cos 2\theta - 4K_4 \cos 4\theta] \qquad (3)$$

where $A_{aa}$ and $A_{cc}$ are second-order and $K_4$ is the fourth-order anisotropy energy constants. E is the exchange constant, S is the sublattice magnetic moment, and θ is the angle between *c*-axis and direction of weak ferromagnetic moment. In TmCrO$_3$, Cr$^{3+}$ spins are in Γ$_2$ (G$_z$, F$_x$) configuration. Here, implementing θ=90° and neglecting the fourth-order term (K<<A), we get

$$\left(\hbar\omega_{qFMR}\right)^2 = \left[\frac{4E}{(2S)^2}\right][(A_{cc} - A_{aa})] \qquad (4)$$

In TmCrO$_3$, Cr$^{3+}$ spins lie in *z*-direction with slight canting above spin reorientation temperature (T$_{SR}$), which implies $A_{cc}$> $A_{aa}$. Below T$_{SR}$, spins move from *c*-axis towards *a*-axis which suggests an increase in $A_{aa}$.[19] As a result ($A_{cc}$ – $A_{aa}$) has higher value above spin-reorientation temperature. This anisotropic shift of spins towards *a*-axis using the ferromagnetic resonance expression of two-sublattice model explains the low temperature dependent frequency shift of $g_1$.

ii) The magnetic ground state energy/frequency of $g_1$ is in lower THz frequency range (extreme end of GHz range) - a behaviour typical to qFMR.[12] With decreasing temperature, this mode moves out of our THz detection range and moves into the GHz -microwave region – particularly known for ferromagnetic resonances. In addition, this mode is robust in applied fields below the spin-reorientation temperature. The fact that this mode shifts towards the higher frequencies with increasing magnetic field is an established magnetic-field tuneable characteristic feature of ferromagnetic resonance.[26] To surmise, all above-mentioned characteristics of $g_1$ mode suggest it to be a quasi-ferromagnetic resonance.

In rare-earth oxides, coexistence of spin excitations and crystal field excitations at different/same THz frequency is possible,[4,7,27] demonstrating an intricate interplay between magnetic and structural elements. Spin excitations, as characterised by collective spin behaviour, manifest in magnetic materials as spin waves or magnons, while crystal field excitations are due to the influence of the local environment on electronic energy levels.

Now we turn our attention to the structure and dynamics of resonant modes marked as $g_2$ (0.56 THz) and $g_3$ (1.25 THz) [Figure 2 (a)]. These modes manifest around 90 K and strengthen with decreasing temperature. Below 30 K, $g_2$ and $g_3$ split into modes denoted by ($g_2^a$, $g_2^b$, $g_2^c$) and ($g_3^a$, $g_3^b$) respectively [Figure 3]. Further down to 10 K, $g_3^a$ and $g_3^b$ splits into two modes ($g_3^{a1}$, $g_3^{a2}$) and ($g_3^{b1}$, $g_3^{b2}$) respectively, which merge back to their original shape at 5 K. The surrounding crystal field, in $TmCrO_3$, lifts the degeneracy of ground state multiplet $^3H_6$ of $Tm^{3+}$ ion [Figure S6 (a), supplemental material]. Consequently, multiple states are generated for crystal-field excitations. The resonances $g_2$ and $g_3$ are assigned to be such low-energy crystal-field excitations for the following reasons: i) the splitting of lowest-lying electronic energy levels of $Tm^{3+}$ ions in the surrounding environment in known to be 2 meV (~ 0.50 THz) and 5 meV (~ 1.22 THz) in $TmFeO_3$, which is an isostructural counterpart of $TmCrO_3$.[28] ii) indirect methods via theoretical modelling on $TmCrO_3$ magnetic susceptibility and magnetization data revealed the existence of crystal-field excitation at ~2 and 5 meV.[29] iii) Inelastic neutron spectroscopy[30] and optical transition studies[31] too have suggested the same. Based on these information and conjectures, we assign two broad modes $g_2$ and $g_3$ in THz absorption spectra appearing at ~90 K as crystal field excitations (CFE).

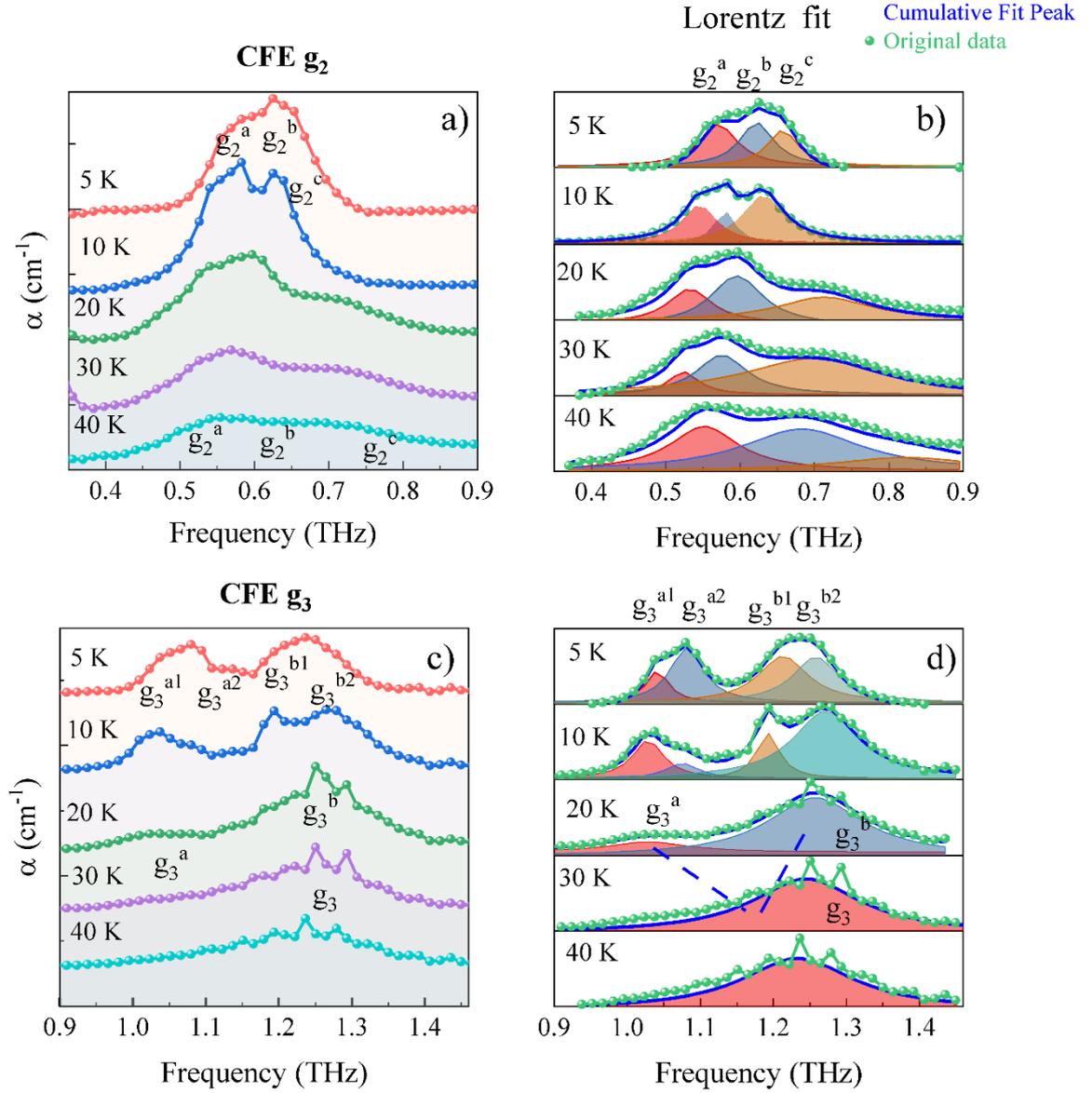

*Figure 3 (a): Temperature dependence of absorption coefficient (α) of $g_2$, (b): Deconvolution of $g_2$ CFE mode using Lorentz peak fit corresponding to different temperatures. (c): Temperature dependence of absorption coefficient (α) of $g_3$, (d): Deconvolution of $g_3$ CFE mode using Lorentz peak fit corresponding to different temperatures. Dashed lines show the splitting of modes. The Curves have been shifted along the Y axis to enhance visibility in 3 (a) and 3 (c).*

Additionally, we report the observation of the fine splitting of crystal field levels (bifurcation/combination of CFE modes) resolved in the current work owing to a higher resolution of the probe tool (0.014 THz=0.0574 meV). Figure 3 (a) presents the evolutions of CFE $g_2$ as temperature is decreased. Below 50 K, $g_2$ splits into finer CFE modes $g_2^a$, $g_2^b$ and $g_2^c$ which grow apart upon further lowering the temperature. However, these modes eventually merge at 5 K. These modes are modelled by Lorentz peak fitting analysis [Figure 3 (b)] depicting a clear presence of three finer CFE in the $g_2$ resonance. In contrast to relatively simple CFE $g_2$, CFE $g_3$ is observed to have much more complex dynamics. The CFE $g_3$ splits into finer CFE $g_3^a$ and $g_3^b$ as temperature is lowered to 20 K [Figure 3 (c)]. These modes further branch off to $g_3^{a1}$ and $g_3^{a2}$, and $g_3^{b1}$ and $g_3^{b2}$ respectively, as the temperature is decreased to 10 K, and then merge back at 5 K. This bifurcation is depicted by the dashed lines in the Lorentz deconvolution of $g_3$ [Figure 3 (d)].

In magnetic materials, the application of magnetic field can cause Zeeman splitting and energy level shifts of CFEs. The Zeeman effect induces line splitting as a consequence of interaction between the magnetic field and CFE magnetic moments. This Zeeman shift is proportional to the strength of applied magnetic field. To gain more insights into field-induced dynamics of crystal-field transitions, magnetic field-dependent THz absorption coefficient at 5 K is depicted in Figure 4.

On the application of magnetic field, $g_2$ shows additional CFE modes along with the existing $g_2^a$, $g_2^b$ and $g_2^c$. The strength of these CFEs increases at higher fields with a gradual shift towards higher frequencies [Figure 4 (a)]. On the other hand, in bifurcated zero-field $g_3$ at 5 K, field-induced modes emerge along with existing CFE modes [Figure 4 (a)]. The magnetic field induced changes are also evident in the different behaviour of the transmitted THz waveforms at 0 T and 5 T measured at 5 K [Figure 4 (b)]. Prominent changes are observed in THz peak position with magnetic field [Inset, Figure 4 (b)]. As magnetic field increases the peak position decreases and has a minimum at 2 T, then slightly increases till 5 T [Inset, Figure 4 (b)].

The magnetic field-induced distinct splitting in these CFE modes is depicted by deconvoluting these modes by Lorentz peak fit, shown in the supplemental material. The bifurcated zero field $g_2$ and $g_3$ modes further split with an increase in magnetic field [Figure S8, supplemental material]. The peak position and strength of these CFEs increase with increasing magnetic field. In $g_3$ mode, as the magnetic field is increased to 5 T, the splitting effect reduces as the number of modes decreases amid the broadening effect, which is so dominant that all split modes appear as a continuum or a single broad mode [Figure 4 (a)]. This broadening suggests field-induced mixing of the crystal field levels. [32] The Zeeman splitting of the crystal field levels also depends on the orientation of the magnetic field. [33,34] In a single crystal, excitation is only possible when a magnetic field is applied along a particular orientation aligning with the direction of crystal field levels. On the contrary, in a polycrystal, the random orientation of the various crystallites causes multiple transitions in the crystal field levels [Figure S6 (b), supplemental material], which can be seen in the THz spectrum as the broad continuum of $g_3$ mode with a magnetic field.

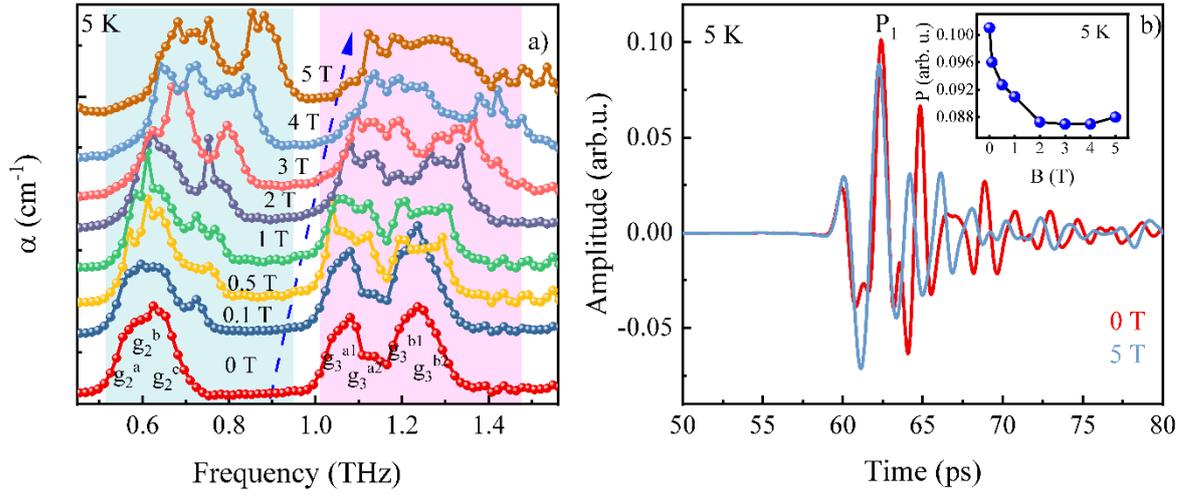

*Figure 4(a): Magnetic field dependence of absorption coefficient (α) of mode $g_2$ and $g_3$ at 5 K. The curves have been shifted along the Y axis to enhance visibility. (b): Typical transmitted THz waveforms at various magnetic field at 5 K. Inset: The variation of THz peak amplitude (P) of peak $P_1$ denoted in THz waveform with magnetic field.*

Identification of fine structure of crystal-field excitations $g_2$ and $g_3$ and their temperature and magnetic-field dependent splitting/mixing using high-resolution THz spectroscopy unveils in greater depth, the structure and dynamics of CFEs in TmCrO$_3$, depicted in Figure 5. Clearly, we observe additional finer levels with the control of temperature and magnetic field. Within the temperature range of 5 K to 50 K, the crystal field levels corresponding to these multiplet of THz excitations modes [Figure 5 (a)], while in magnetic field of up to 5 T, a richer and finer resolved CFE multiplet at 5 K and 20 K is plotted in Figure 5 (b). A comparison of these two sets of THz CFE multiplets with previously inelastic neutron scattering studies [Figure 5 (c)] clearly showcases a significantly enhanced understanding of the crystal field structure of TmCrO$_3$ in particular, and chromites in general, that can be achieved using a novel optical tool of high resolution such as THz spectroscopy.

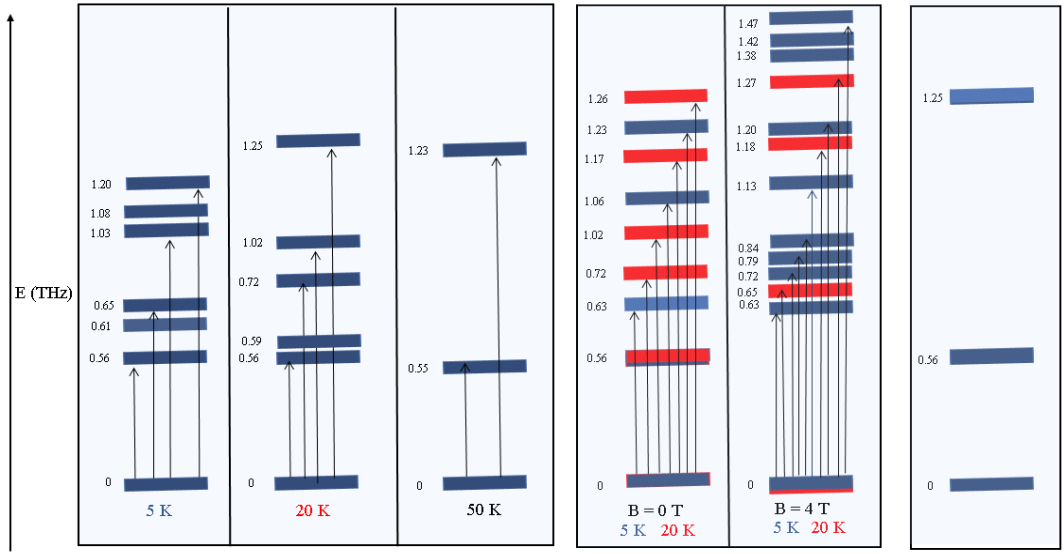

*Figure 5 (a): Crystal field transitions in electronic energy levels of $Tm^{3+}$ in $TmCrO_3$ explored in our THz study at 5 K, 20 K and 50 K. (b): Transitions in $Tm^{3+}$ levels induced by magnetic field at 0 T and 4 T corresponds to 5 K and 20 K. (c): Crystal field levels discovered in previous inelastic neutron scattering studies at 4 K.*

Recent two-dimensional (2D) terahertz spectroscopic studies on $YFeO_3$ have realized the non-linear magnon-magnon interactions, revealing the addition and subtraction of two magnon excitation frequencies.[16] This magnon algebra at THz frequencies has opened a new research area of noncollinear magnets-based magnon spintronics, which reaps the benefits of ultrafast response with negligible dissipative effects.[16,17] $TmCrO_3$ is enriched with a low frequency magnon excitation and numerous crystal field excitations [Figure 5 (a)], associated with non-trivial magnetic interactions across various transition temperatures. This coexistence of closely spaced excitations spanning a narrow THz frequency range is infrequent in rare earth orthochromites and is a necessary condition for 2D THz spectroscopy,[16] for realizing hybrid crystal field and magnon based computation for futuristic technology.

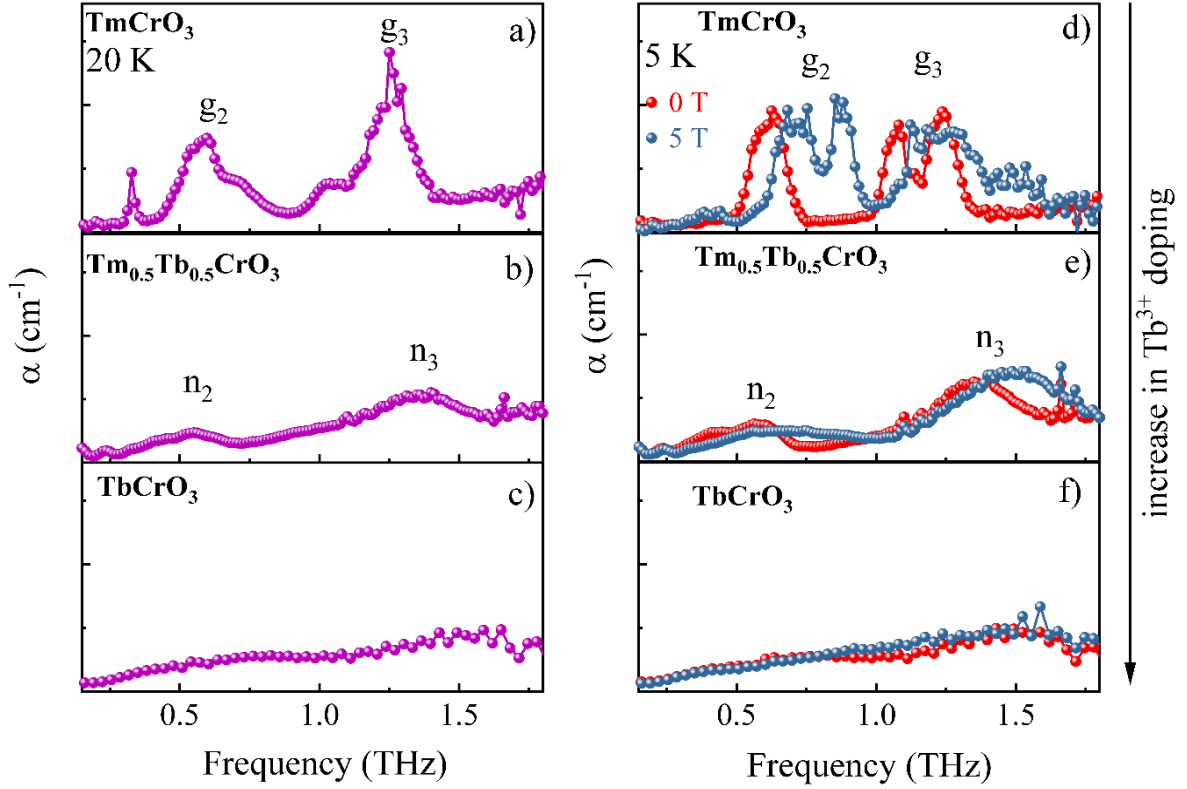

*Figure 6 (a-f): Comparison of absorption coefficient (α) of TmCrO$_3$, Tm$_{0.5}$Tb$_{0.5}$CrO$_3$, TbCrO$_3$. (a-c): at 20 K in zero magnetic field. (d-f): at 5 K with 0 T and 5 T.*

The origin of observed spin-excitations is attributed to the magnetic interaction between Cr and Tm subsystems and that of CFEs lies in electrostatic potential experienced by Tm$^{3+}$ ion due to surrounding crystal field. The rare-earth element plays a pivotal role in determining the spin-excitations and the electronic transitions in rare-earth chromates. Tm$^{3+}$ based rare-earth chromates is known to possess a plethora of temperature-induced magnetic transitions from antiferromagnetism, compensation, spin-reorientation, to rare-earth ordering (< 2 K), while other high magnetic moment RE based chromates, ex. Tb$^{3+}$, possess only antiferromagnetic and rare-earth ordering (at 5 K). [3,35] Substitution of Tb$^{3+}$ at the Tm$^{3+}$ site can further help bring out the role of single-ion anisotropy in multiple magnetic transitions and its underlying effects on spin and crystal-field excitations. To investigate this, we systematically substituted Tb$^{3+}$ at the Tm$^{3+}$ site, in the resultant compositions TmCrO$_3$, Tm$_{0.5}$Tb$_{0.5}$CrO$_3$, and TbCrO$_3$ it was observed that the T$_N$ increases from 125 K to 140 K to 157 K, respectively [Figure S5 (a), (b), (c) of supplemental material].

A detailed analysis of the THz measurement of $Tm_{0.5}Tb_{0.5}CrO_3$ is given in supplemental material [Figure S9], a comparison of the THz absorption spectra of $Tb^{3+}$ substituted $TmCrO_3$ shows suppression of the broad crystal field resonant modes as $Tb^{3+}$ concentration increases [Figure 6 (a-f)]. In $TmCrO_3$, magnon excitation $g_1$ and bifurcated CFE modes of $g_2$ and $g_3$ appear at 20 K [Figure 6 (a)]. In $Tm_{0.5}Tb_{0.5}CrO_3$, the strength of CFEs (denoted by $n_2$ and $n_3$) is lowered along with reduced splitting. And these modes completely disappear for $TbCrO_3$ [Figure 6 (b) and (c)]. Magnetic-field dependence of these modes shows remarkable changes at 5 K [Figure 4]. In $TmCrO_3$, magnon excitation and CFEs split are observed, which is accompanied by shifting towards higher frequencies at 5 T [Figure 6 (d)]. In $Tm_{0.5}Tb_{0.5}CrO_3$, the magnetic field blue shifts those two CFE modes, which were observed at zero magnetic field [Figure 6 (e)], however, in $TbCrO_3$ no significant changes were observed with magnetic field [Figure 6 (f)].

The effect of $Tb^{3+}$ substitution on the CFEs can be understood in the following context. The ionic radii of $Tb^{3+}$ is greater than that of the $Tm^{3+}$ ion ($r_{Tb} > r_{Tm}$), which decreases orthorhombic distortion accompanied by an increase in the $Cr^{3+}$-$O^{2-}$-$Cr^{3+}$ bond angle.[36] The resulting tilting and rotation of oxygen octahedra responsible for crystalline environment at $Tm^{3+}$ site, subsides the CFEs with $Tb^{3+}$ doping.[37] The anisotropy of $Cr^{3+}$ sublattice is larger in $TmCrO_3$ compared to $TbCrO_3$ because of increasing $Cr^{3+}$-$O^{2-}$-$Cr^{3+}$ bond angle with increasing ionic radii,[38] thereby shifting the CFE modes to higher energies. This corroborates well with $TbCrO_3$ inelastic neutron scattering where the crystal field excitations are observed at ~ 20 meV (= 4.8 THz).[30] While the disappearance of spin excitations is associated with increase in $Cr^{3+}$-$O^{2-}$-$Cr^{3+}$ bond angle resulting in reduction of magnetic interaction energy of $Tm^{3+}$- $Cr^{3+}$ ions.

Current THz investigation highlighted the fundamental and technological significance of noncollinear magnet $TmCrO_3$, prompting renewed interest in the understudied family of orthochromites. These discoveries encourage the exploration of magnetic materials with intricate magnetic ground states that possess more than one magnon or other hybrid modes at THz frequencies such that interaction between them can create new magnon frequency states relevant to quantum information processing.

In the rapidly advancing field of computation technology, quantum computation uses coherent states represented by a superposition of non-trivial quantum states (|0> & |1>), forming a qubit [39] where its execution depends on the superposition and entanglement of the quantum states.[40] The crystal field levels in rare-earth-based systems, being static quantum states, allow for it to possibly act as a qubit state [39] and have advantages over other two-state quantum systems such as trapped ion, superconducting qubits, etc., owing to its longer coherence time and multiple spins created by incorporating nuclear spins with electronic spins via hyperfine interactions.[41] Although the single platform qubits have made progress in quantum information technologies, there are increasing decoherence limits associated with scaling up these devices. Creating hybrid platforms involving qubits' interaction with other quantum states can offer new ways to uncharted quantum regions. [42] Distinct hybrid systems have been designed to perform quantum operations, [43-45] resolving the limitations of single-platform qubit systems and expanding upon the existing research in quantum computation.

Based on our results, we propose a "proof-of-concept application" demonstrating the practical conceptualization of the Qubit-magnon hybrid system [Figure 7]. THz radiation interacts with orthorhombically distorted $TmCrO_3$ [Figure 7 (a)], resulting in the observation of crystal fields and magnon excitations [Figure 7 (b)]. The magnon excitation originates from the interaction of Cr and Tm sublattices, and crystal field excitation corresponds to the transitions in ground state multiplet $^3H_6$ of $Tm^{3+}$ ion. These closely spaced excitations, which show temperature-dependent modulation of their peak frequencies, show a strong possibility of intrinsic mutual interaction in the THz energy range which can be probed with 2D THz spectroscopy [16] [Figure 7 (c)]. These type of interactions in heterogeneous excitations is achieved in artificially engineering structures with the aim of simultaneously harnessing the advantages of different quantum systems in a single device. [43] Remarkably, the coexistence of magnon and crystal field excitations in a single material $TmCrO_3$, is observed with this novel THz technique, which is an intrinsic feature of the compound. These findings encourage the further exploration of materials exhibiting such intrinsic features which has potential to execute quantum operations.

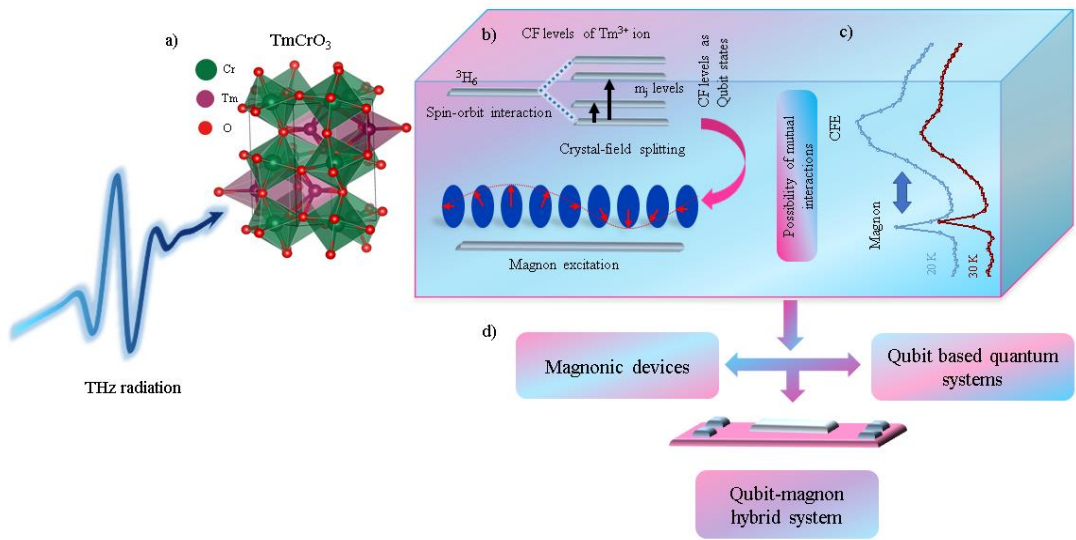

*Figure 7: Schematic illustrating (a) interaction of terahertz (THz) radiation with orthorhombically distorted $TmCrO_3$. (b) This interaction results simultaneous presence of magnon and crystal field transitions in energy levels of $Tm^{3+}$ ion in THz spectrum of $TmCrO_3$. These different levels of $Tm^{3+}$ can serve as qubit states used for computing devices. (c) In THz spectrum these excitations are closely spaced, thereby creating the*

*possibility of mutual interactions in these heterogeneous collective excitation modes. (d) These results open the door for designing Qubit- magnon hybrid system by integrating qubit-based quantum system with magnonic devices.*

Magnonic device use coherent manipulation of magnons to transfer information, while qubit systems use the superposition of quantum states to transfer information. Integrating these devices in one platform results in a Qubit-*magnon hybrid system* [Figure 7 (d)]. In this platform, magnons can mediate the qubits' interactions, decreasing the decoherence over larger distances. In addition, by tuning this interaction between magnons and qubits, the quantum states can be controlled in a precise manner.

Thus, the evolution of crystal field and magnon excitations and the possibility of their mutual interactions in THz regime can add new functionality in quantum computation devices that can work on an ultrafast timescale and with ultra-low power dissipation, thereby boosting the field of THz quantum computation.

## IV. CONCLUSION

This study on the THz low-energy magnetic ground state of orthochromite, $TmCrO_3$, shows a myriad of spin-wave and crystal-field based excitations depicting the interplay of the magnetic exchange interaction within and between Tm and Cr subsystems. The temperature and magnetic field dynamics of CFEs and spin excitations adds a new dimension in understanding the electronic structure of this orthochromite. The prominent role of rare-earth sites in $RE^{3+}$-$Cr^{3+}$ interaction and its effect on spin and CFEs was further understood by the substitution of $Tb^{3+}$ into the $Tm^{3+}$ matrix. A THz all-optical control of a broad range of magnetic phenomena not only holds importance in the fundamental understanding of orthochromites magnetism but investigation via display of multitude of closely spaced low-energy excitation modes brings out possible technological relevance such as in crystal field hybrid spin-wave computing in the ambit of THz magnon spintronics. Finally, the advantage of averaged effect in a polycrystalline sample can be efficiently explored for faster screening of materials thereby boosting the field of THz magnetism.

**Notes**

The authors declare no competing financial interest.

## ACKNOWLEDGEMENTS

D.S.R. acknowledges the Science and Engineering Research Board (SERB), Department of Science and Technology, New Delhi, for financial support under research Project No. CRG/2020/002338. G.D. acknowledges support from the funding agency, University Grant Commission (UGC) of Government of India for SRF fellowship. B.S.M. thanks Prime Minister

Research Fellowship (PMRF; 0401968) funding agency, Ministry of Education. The authors acknowledge Central Instrumentation Facility at IISER Bhopal for SQUID- VSM and XRD facilities.# References

(1) Yoshii, K. Magnetization Reversal in TmCrO3. *Mater. Res. Bull.* **2012**, *47*, 3243.

(2) Rajeswaran, B.; Khomskii, D. I.; Zvezdin, A. K.; Rao, C. N. R.; Sundaresan, *A.* Field-Induced Polar Order at the Néel Temperature of Chromium in Rare-Earth Orthochromites*:* Interplay of Rare-Earth and Cr Magnetism. *Phys. Rev. B* **2012** 86, 214409.

(3) Yin, L. H.; Yang, J.; Tong, P.; Luo, X.; Park, C. B.; Shin, K. W.; Song, W. H.; Dai, J. M.; Kim, K. H.; Zhu, X. B.; Sun, Y. P. Role of Rare Earth Ions in the Magnetic, Magnetocaloric and Magnetoelectric Properties of RCrO3 (R = Dy, Nd, Tb, Er) Crystals. *J. Mater. Chem. C* **2016**, *4*, 11198.

(4) Fabrèges, X.; Petit, S.; Brubach, J. B.; Roy, P.; Deutsch, M.; Ivanov, A.; Pinsard-Gaudart, L.; Simonet, V.; Ballou, R.; De Brion, S. Interplay between Spin Dynamics and Crystal Field in the Multiferroic Compound HoMnO3. *Phys. Rev. B* **2019**, *100*, 094437.

(5) Preethi Meher, K. R. S.; Wahl, A.; Maignan, A.; Martin, C.;Lebedev, O. I. Observation of electric polarization reversal and magnetodielectric effect in orthochromites: A comparison betweenLuCrO3and ErCrO3.*Phys. Rev. B* **2014**,89, 144401.

(6) Shen, H.; Xian, Q.; Xie, T.; Wu, A.; Wang, M.; Xu, J.; Jia, R.; Kalashnikova, A. M. Modulation of Magnetic Transitions in SmFeO3 Single Crystal by Pr3+ Substitution. *J. Magn. Magn. Mater.* **2018**, *466*, 81.

(7) Zhang, K.; Xu, K.; Liu, X.; Zhang, Z.; Jin, Z.; Lin, X.; Li, B.; Cao, S.; Ma, G. Resolving the Spin Reorientation and Crystal-Field Transitions in TmFeO3 with Terahertz Transient. *Sci. Rep.* **2016**, *6*, 23648.18

# Terahertz crystal-field transitions and quasi ferromagnetic magnon excitations in a noncollinear magnet for hybrid spin-wave computation

Gaurav Dubey[1], Brijesh Singh Mehra[1], Sanjeev Kumar[1], Ayyappan Shyam[1], Karan Datt Sharma[1], Megha Vagadia[1], Dhanvir Singh Rana[1]*

[1]Indian Institute of Science Education and Research Bhopal, 462066, India

*dsrana@iiserb.ac.in

## S1: Structural Characterization

$TmCrO_3$, $Tm_{0.5}Tb_{0.5}CrO_3$ and $TbCrO_3$ polycrystalline samples were prepared by standard solid state reaction method. The powders $Tm_2O_3$, $Tb_2O_3$ and $Cr_2O_3$ are weighed in the proper stoichiometric ratio, mixed, grounded using a mortar pestle, and calcinated in the furnace at 1100°C for 36 hours. The calcinated powders were reground, pelletized in 10 mm diameter and thickness 1 mm, and sintered at 1450°C for 36 hours.

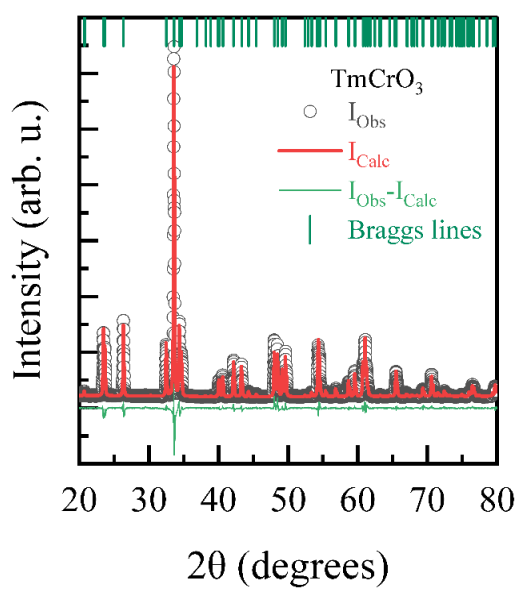 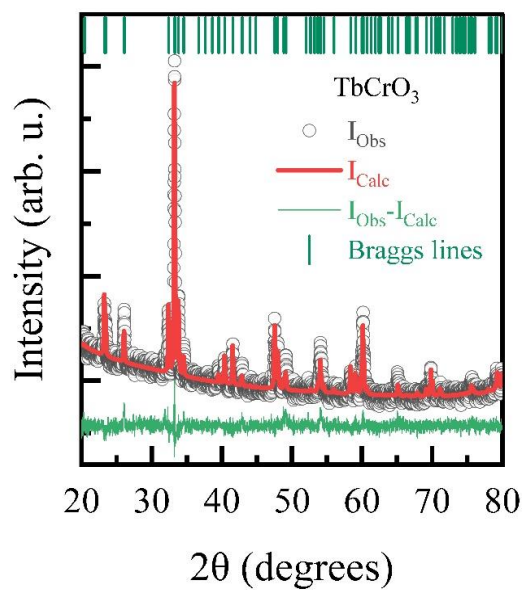

*Figure S1: XRD Rietveld refinement of $TmCrO_3$.*   *Figure S2: XRD Rietveld refinement of $TbCrO_3$*

The Sample was examined by X-ray diffraction method using Cu $K_\alpha$ radiation. To confirm phase purity, Rietveld refinement of XRD data of powder sample is done using Profex software. XRD data of TmCrO$_3$ reveals that the sample is in single phase with no observed impurity [Figure S1]. The obtained lattice parameters are a = 5.214 Å, b = 5.506 Å, c = 7.507 Å which are close to previously reported values.[1] The value of goodness of fitting (GOF) is 1.65. Similarly, for TbCrO$_3$, the XRD data confirms the single-phase nature of polycrystalline sample with no observed impurity and calculated lattice parameters are a = 5.296 Å, b = 5.518 Å, c = 7.578 Å [Figure S2]. The value of GOF is 1.22. These lattice parameters with the previously reported values.[2]

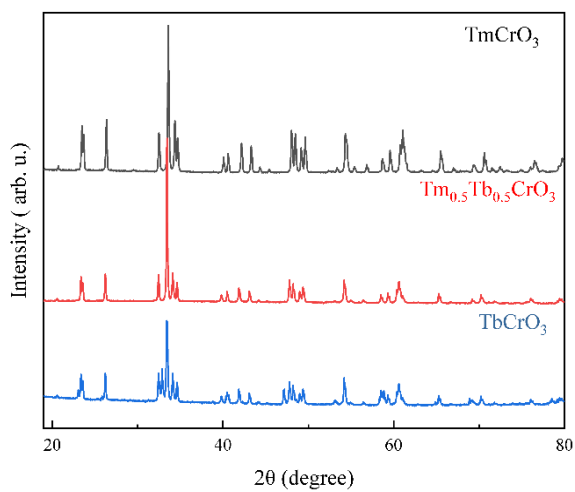

*Figure S3: XRD comparison of TmCrO$_3$, Tm$_{0.5}$Tb$_{0.5}$CrO$_3$ and TbCrO$_3$*

Additionally, in powder XRD spectrum of Tm$_{0.5}$Tb$_{0.5}$CrO$_3$, peak features lie in between TmCrO$_3$ and TbCrO$_3$ which further confirms successful formation of Tm$_{0.5}$Tb$_{0.5}$CrO$_3$ [Figure S3].

## S2: Elemental Characterization: EDAX

To check the chemical composition, we have done energy dispersive x- ray diffraction (EDAX) in Tm$_{0.5}$Tb$_{0.5}$CrO$_3$ sample. EDAX measurements show that in Tm$_{0.5}$Tb$_{0.5}$CrO$_3$ the constituent elements (Tm and Tb) are present in equal composition [Table I].

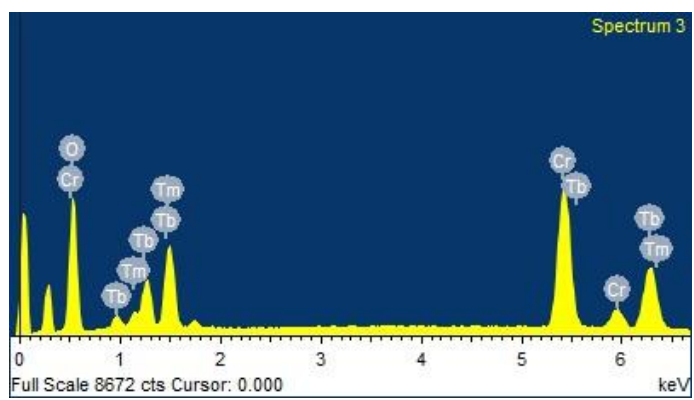

*Figure S4: EDAX pattern for Tm$_{0.5}$Tb$_{0.5}$CrO$_3$.*

Table I: Elemental composition in $Tm_{0.5}Tb_{0.5}CrO_3$

| Spot | Atomic percentage of elements | | | |
| --- | --- | --- | --- | --- |
| | Tm | Tb | Cr | O |
| 1. | 10.39 | 9.62 | 18.52 | 61.47 |
| 2. | 9.60 | 8.84 | 17.29 | 64.27 |
| 3. | 12.48 | 11.60 | 21.93 | 53.99 |
| Average | 10.82 | 10.02 | 19.24 | 59.91 |

Figure S4 shows the composition analysis of elements present in $Tm_{0.5}Tb_{0.5}CrO_3$. The atomic percentage of Tm and Tb are found to be ~ 0.54 and ~ 0.50 respectively, which depict the successful formation of $Tm_{0.5}Tb_{0.5}CrO_3$ [Table I]. To analyse their magnetic behaviour, we have performed magnetic measurements on these samples.

### S3: Magnetic Characterization

The magnetic properties of these samples are studied with temperature and magnetic field dependent DC magnetization using SQUID (Quantum Design) magnetometer in the VSM mode in the temperature range 2-300 K and in the presence of magnetic field up to 9 T/-9 T. The magnetic behaviour of these samples is depicted in Figure S5. In $TmCrO_3$ Neel temperature ($T_N$) is ~125 K [Figure S5 (b)]. In $Tm_{0.5}Tb_{0.5}CrO_3$ the Neel temperature ($T_N$) corresponding to $Cr^{3+}$ ordering is ~ 140 K [Figure S5 (b)]. The effective magnetic moment obtained by Curie Weiss fit on inverse susceptibility curve of $Tm_{0.5}Tb_{0.5}CrO_3$ is 9.42 $\mu_B$/f.u [Inset, Figure S5 (b)]. In $TbCrO_3$, the Neel temperature ($T_N$) corresponding to $Cr^{3+}$ and $Tb^{3+}$ ordering is ($T_{N1}$) ~ 157 K and ($T_{N2}$) ~ 5 K respectively [Figure S5 (c)]. Curie Weiss fit on inverse susceptibility curve of $TbCrO_3$ [Inset, Figure S5 (c)] gives value of effective magnetic moment 10.68 $\mu_B$/f.u. which matches with the previously reported values. [2] The effective magnetic moment of $Tm_{0.5}Tb_{0.5}CrO_3$ lies in between $TmCrO_3$ and $TbCrO_3$.

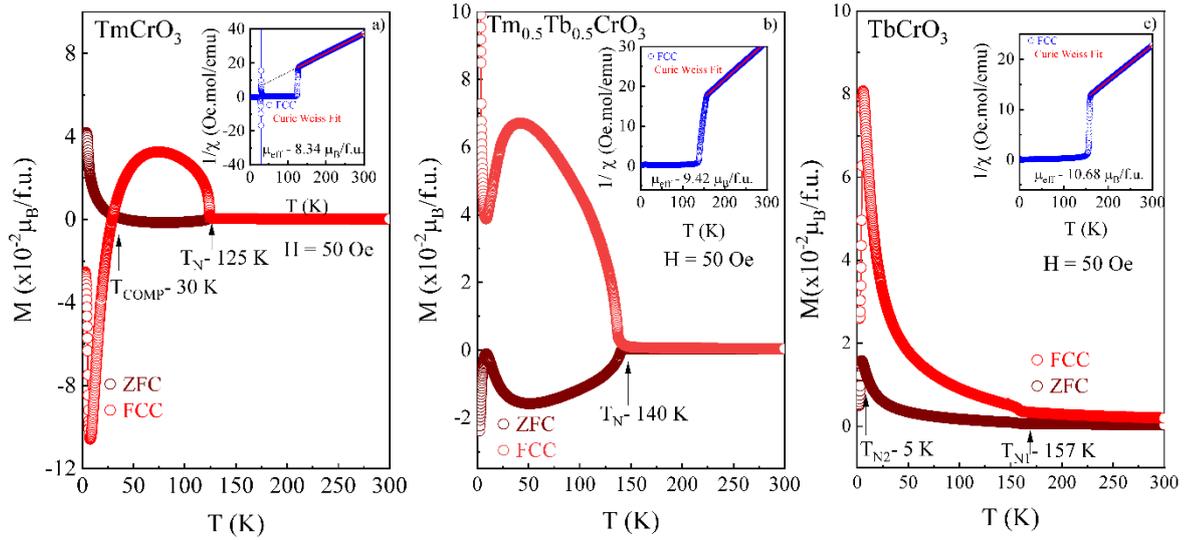

*Figure S5: Magnetization as a function of temperature in ZFC and FCC cycle of under an applied magnetic field of 50 Oe. Inset: Reciprocal of magnetic susceptibility as a function of temperature, red lines show the Curie Weiss fit on inverse susceptibility curve (a) TmCrO$_3$ (b) Tm$_{0.5}$Tb$_{0.5}$CrO$_3$. (c) TbCrO$_3$.*

## S4: Crystal field transitions

In TmCrO$_3$, the crystal field surrounding the Tm$^{3+}$ ion splits ground state multiplet created due to spin-orbit coupling into nondegenerate states [Figure S6]. These levels are denoted as m$_J$ levels. The energy of these crystal field levels varies when the magnetic field is applied due to the Zeeman effect. The lowest energy levels are at 2.54 meV (~ 0.50 THz) and 5.12 meV (~ 1.23 THz),[3] corroborating inelastic neutron scattering.[4] In a single crystal, the transition in the different levels occurs when the magnetic field is applied along a particular direction. In contrast, in the polycrystalline sample, the crystallites are oriented in different directions [Figure S6 (b)]. Applying a magnetic field in different direction causes different Zeeman splitting;[5] average behaviour of transitions in these Zeeman split levels is obtained in a magnetic field.

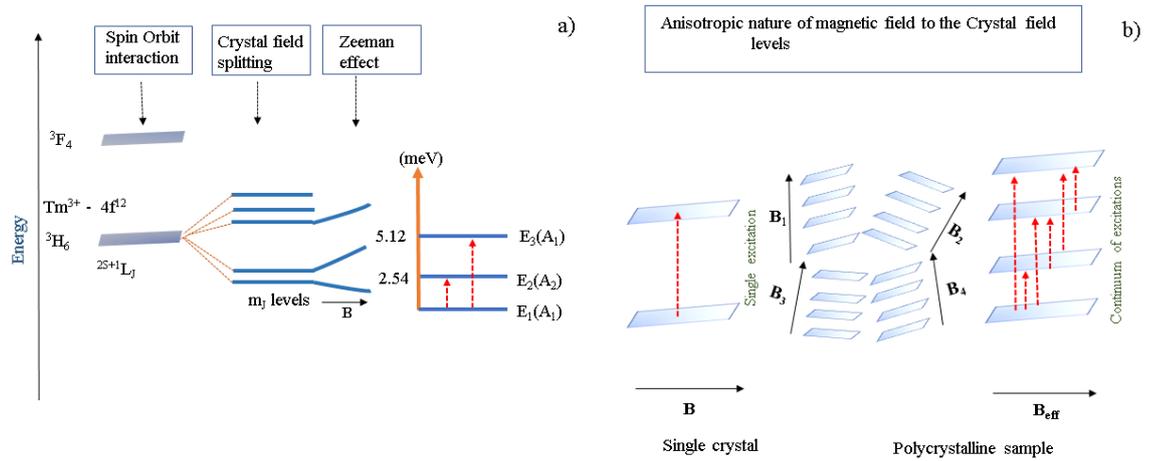

*Figure S6 (a): Crystal field splitting of $Tm^{3+}$ ion in $TmCrO_3$. (b): Anisotropic nature of magnetic field to the crystal field levels.*

## S5: Magneto-THz time-domain experiment:

### S5-1: Analysis of THz waveform of $TmCrO_3$:

To investigate the low energy dynamics, magneto-terahertz (THz) measurements were performed on these samples. In THz waveform [Figure 1 (b)], the peak position ($P_1$) shifts in time at 60 K (shown by arrow) with change in temperature [Figure S7 (a)]. In THz time domain waveform, there is transfer of normalised temporal weight from lower time scale ($t_1$-$t_2$) (50-64 ps) to higher time scale ($t_2$-$t_3$) (64-80 ps). It increases with decrease in temperature from 60 K (~ 30 %) to 30 K (~ 500 %) to 20 K (~ 700 %) [Figure S7 (b)].

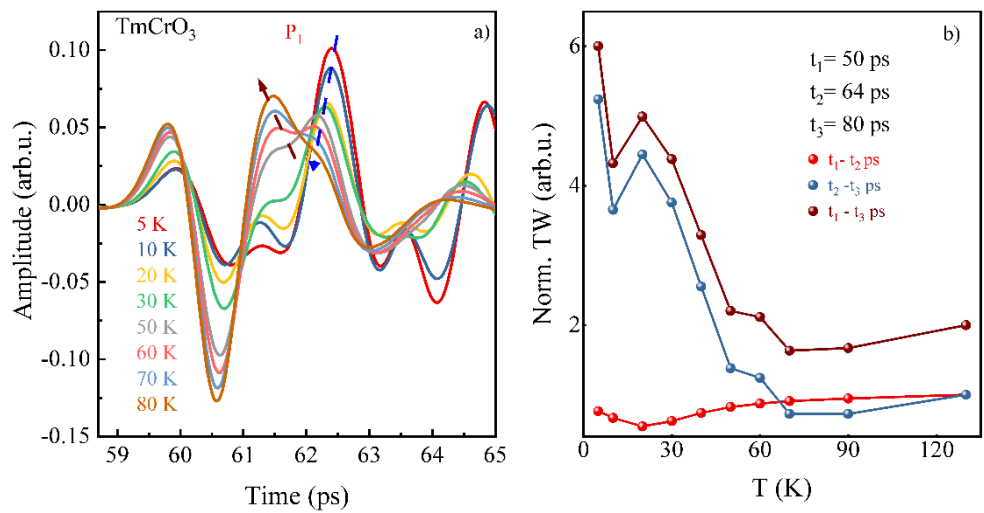

*Figure S7 (a): THz waveform of $TmCrO_3$ depicting change in peak position $P_1$ with change in temperature. (b): Normalized temporal weight (Norm. TW) transfer as a function of temperature variation of $TmCrO_3$.*

**S5-2: THz Absorption of magnetic field dependence of $g_2$ and $g_3$ of TmCrO$_3$**

In TmCrO$_3$, at 5 K CFEs $g_2$ and $g_3$ spilt and shift towards higher frequencies with increase in magnetic field. This magnetic field induced splitting is shown by Lorentz peak fit to the CFE modes [Figure S8]. As can be seen in Lorentz peak deconvolution [Figure S8 (a)], with increasing the magnetic field the modes $g_2^a$, $g_2^b$ and $g_2^c$ further split into multiple modes with shifting towards higher frequencies. The magnetic field dynamics of $g_3$ is more vibrant. The modes $g_3^{a1}$, $g_3^{a2}$, $g_3^{b1}$ and $g_3^{b2}$ split further on the application of magnetic field with slight shifting towards higher frequencies. The Lorentz peak deconvolution shows the splitting of these modes on the application of magnetic field [Figure S8 (b)]. The number of modes increase with broadening on the application of magnetic field and at 5 T these modes make a broad continuum.

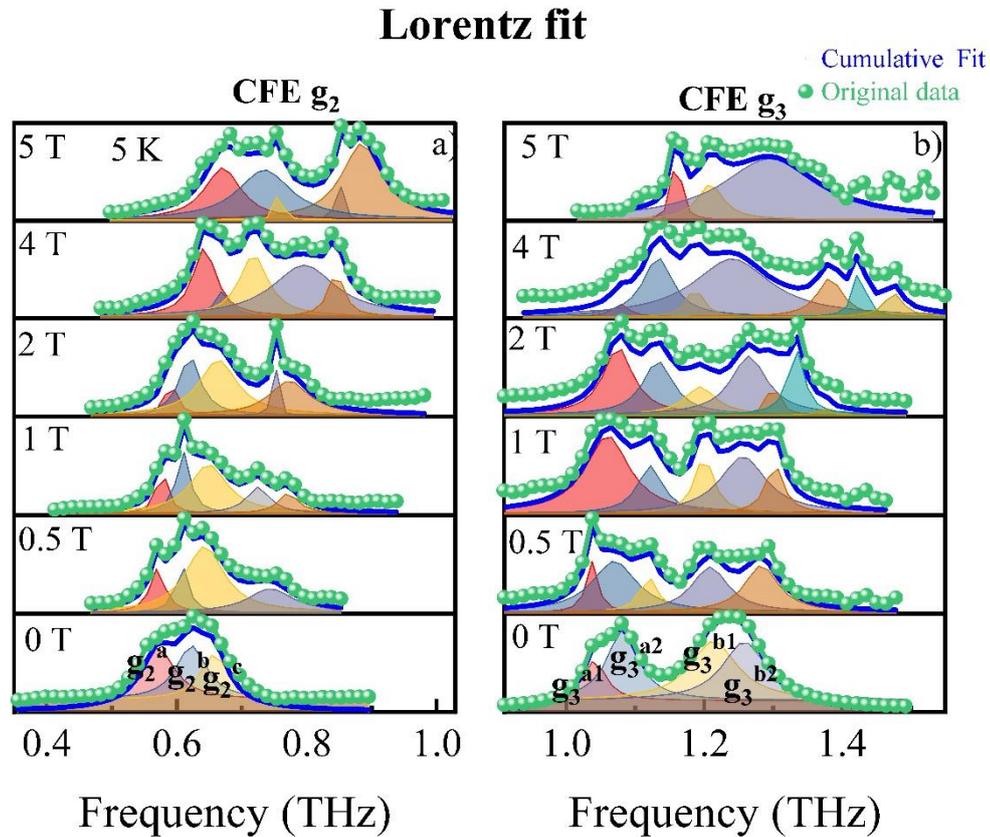

*Figure S8: Lorentz peak deconvolution of magnetic field dependence of CFEs (a) $g_2$ and (b) $g_3$ of TmCrO$_3$ at 5 K.*

**S5-3: THz Absorption spectrum of Tm$_{0.5}$Tb$_{0.5}$CrO$_3$**

In magneto-terahertz investigation of Tm$_{0.5}$Tb$_{0.5}$CrO$_3$, absorption spectra show two broad resonant modes ($n_2$ at 0.56 THz and $n_3$ at 1.30 THz) appearing below 70 K which enhances with the decrease in temperature [Figure S9 (a)]. In accordance with TmCrO$_3$, these are the CFEs, in which CFE $n_2$ further splits into $n_2^a$, $n_2^b$ with the decrease in temperature whereas the higher-frequency CFE doesn't, as depicted more in the deconvolution [Figure S9 (b)]. The area under the curve (A) for bifurcated $n_2$ modes is depicted in Figure S9 (c). Area of $n_2^b$ increases accompanied by slight decrease in $n_2^a$. With the magnetic field both CFE shifts to higher frequency, however $n_2$ suppresses and $n_3$ enhances [Figure S9 (d)]. The behaviour of modes $n_2$ and $n_3$ in $Tm_{0.5}Tb_{0.5}CrO_3$ resembles the CFEs observed in $TmCrO_3$, albeit with lower strength. Furthermore, the magnetic excitation was absent in $Tm_{0.5}Tb_{0.5}CrO_3$. Thus, in $Tm_{0.5}Tb_{0.5}CrO_3$, we did not observe a robust excitation with temperature and magnetic field control in comparison with $TmCrO_3$. At 5 K, in absence of magnetic field, crystal field excitations (CFEs) are reducing in number and strength and shift towards higher frequencies with increase in $Tb^{3+}$ doping [Figure 6 (d-f)]. While application of magnetic field in $TmCrO_3$ causes both splitting and shifting in the spectra [Figure 6 (d)], for $Tm_{0.5}Tb_{0.5}CrO_3$, it only results in the red shift without any splitting in resonance modes [Figure S9 (d)].

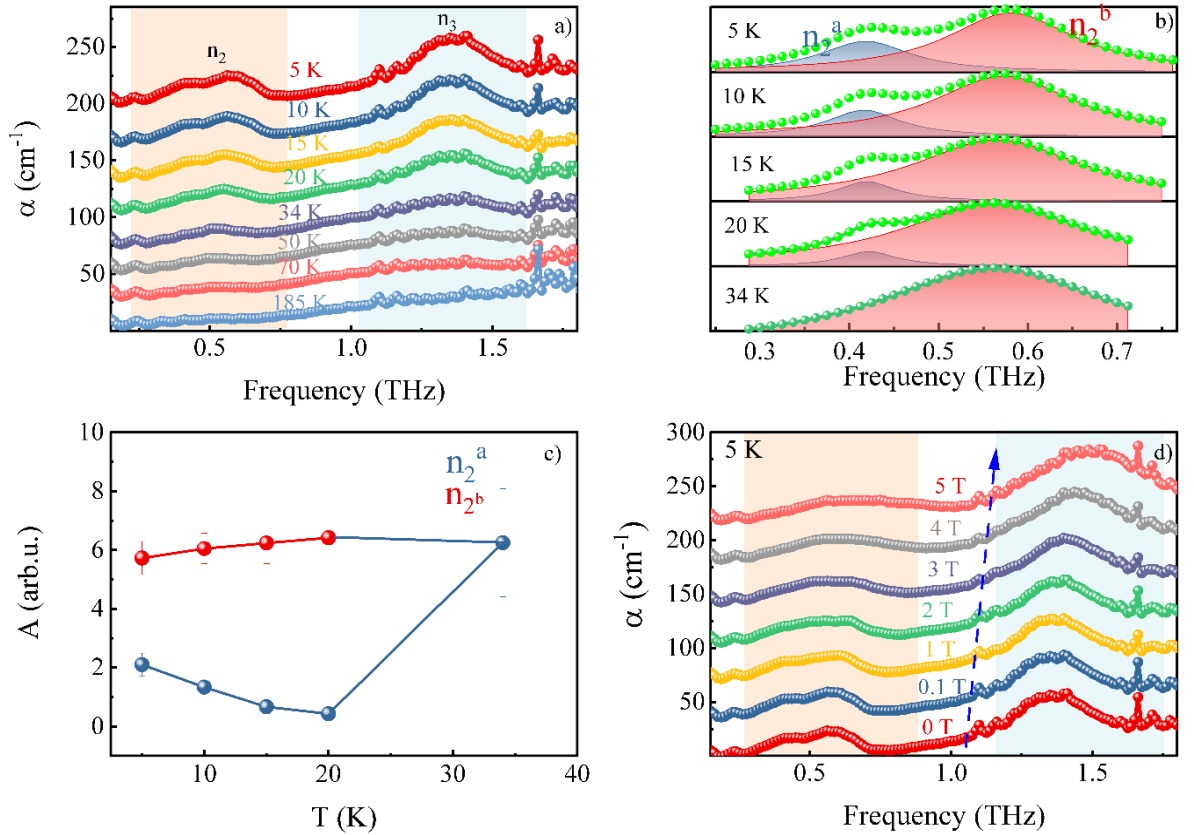

*Figure S9: Absorption coefficient (α) as a function of THz frequency corresponding to different temperatures for $Tm_{0.5}Tb_{0.5}CrO_3$. (b) Deconvolution of $n_2$ peak using Lorentz peak fit at some selected temperatures. (c) Distribution of area under $n_2$ absorption curves (A) in the modes $n_2^a$, $n_2^b$ as a function of temperature (d) Magnetic field variation of absorption coefficient (α) as a function of THz frequency at 5 K. The curves have been shifted along Y axis for better view. Arrow indicates direction of frequency shift with increase in magnetic field.*